\documentclass[twocolumn,times]{aastex63}
\usepackage{amsmath}
\usepackage{overpic}
\usepackage{rotating} 
\makeatletter
\let\splitbox\relax
\makeatother
\usepackage{tabularx} 
\usepackage{amsmath} 
\usepackage{booktabs}
\usepackage{longtable}%
\usepackage{makecell}
\usepackage{enumerate}
\usepackage{adjustbox}
\usepackage{threeparttable}
\usepackage{afterpage}
\usepackage{tipa}
\makeatletter

\newcommand{\Rmnum}[1]{\expandafter\@slowromancap\romannumeral #1@}
\makeatother

\setwatermarkfontsize{1.5in}

\shortauthors{Wang ET AL.}
\begin{document}


\title{Unveil A Peculiar Light Curve Pattern of Magnetar Burst with GECAM observations of SGR J1935+2154}


\correspondingauthor{Shao-Lin Xiong}
\email{xiongsl@ihep.ac.cn}

 \author[0009-0008-5068-3504]{Yue Wang}
 \affil{State Key Laboratory of Particle Astrophysics, Institute of High Energy Physics, Chinese Academy of Sciences, Beijing 100049, China}
 \affil{University of Chinese Academy of Sciences, Chinese Academy of Sciences, Beijing 100049, China}

 \author[0009-0008-8053-2985]{Chen-Wei Wang}
 \affil{State Key Laboratory of Particle Astrophysics, Institute of High Energy Physics, Chinese Academy of Sciences, Beijing 100049, China}
 \affil{University of Chinese Academy of Sciences, Chinese Academy of Sciences, Beijing 100049, China}
 
 \author[0000-0002-4771-7653]{Shao-Lin Xiong*}
 \affil{State Key Laboratory of Particle Astrophysics, Institute of High Energy Physics, Chinese Academy of Sciences, Beijing 100049, China}

\author[0000-0003-2957-2806]{Shuo Xiao}
\affil{Guizhou Provincial Key Laboratory of Radio Astronomy and Data Processing, 
 Guizhou Normal University, Guiyang 550001, People’s Republic of China}
\affil{School of Physics and Electronic Science, Guizhou Normal University, Guiyang 550001, People’s Republic of China}

\author[0000-0001-5348-7033]{Yan-Qiu Zhang}
\affil{State Key Laboratory of Particle Astrophysics, Institute of High Energy Physics, Chinese Academy of Sciences, Beijing 100049, China}
\affil{University of Chinese Academy of Sciences, Chinese Academy of Sciences, Beijing 100049, China}
 
\author[0000-0001-9217-7070]{Sheng-Lun Xie}
\affil{State Key Laboratory of Particle Astrophysics, Institute of High Energy Physics, Chinese Academy of Sciences, Beijing 100049, China}
\affil{Institute of Astrophysics, Central China Normal University, Wuhan 430079, China}

\author[0000-0002-0633-5325]{Lin Lin}
\affil{Department of Astronomy, Beijing Normal University, Beijing 100875, People’s Republic of China}

\author[0000-0001-6374-8313]{Yuan-Pei Yang}
\affil{South-Western Institute for Astronomy Research, Yunnan University, Kunming 650500, China}

\author{Hao-Xuan Guo}
\affil{State Key Laboratory of Particle Astrophysics, Institute of High Energy Physics, Chinese Academy of Sciences, Beijing 100049, China}
\affil{Department of Nuclear Science and Technology, School of Energy and Power Engineering, Xi'an Jiaotong University, Xi'an, China}

\author[0000-0002-6540-2372]{Ce Cai}
\affil{College of Physics and Hebei Key Laboratory of Photophysics Research and Application, Hebei Normal University, Shijiazhuang, Hebei 050024, China}

\author{Yue Huang}
\affil{State Key Laboratory of Particle Astrophysics, Institute of High Energy Physics, Chinese Academy of Sciences, Beijing 100049, China}

\author[0000-0001-5798-4491]{Cheng-Kui Li}
\affil{State Key Laboratory of Particle Astrophysics, Institute of High Energy Physics, Chinese Academy of Sciences, Beijing 100049, China}

\author[0000-0002-0238-834X]{Bing Li}
\affil{State Key Laboratory of Particle Astrophysics, Institute of High Energy Physics, Chinese Academy of Sciences, Beijing 100049, China}

\author[0000-0003-4585-589X]{Xiao-Bo Li}
\affil{State Key Laboratory of Particle Astrophysics, Institute of High Energy Physics, Chinese Academy of Sciences, Beijing 100049, China}
 
\author[0009-0004-1887-4686]{Jia-Cong Liu}
\affil{State Key Laboratory of Particle Astrophysics, Institute of High Energy Physics, Chinese Academy of Sciences, Beijing 100049, China}
\affil{University of Chinese Academy of Sciences, Chinese Academy of Sciences, Beijing 100049, China}

\author[0000-0002-2032-2440]{Xiang Ma}
\affil{State Key Laboratory of Particle Astrophysics, Institute of High Energy Physics, Chinese Academy of Sciences, Beijing 100049, China}
\affil{University of Chinese Academy of Sciences, Chinese Academy of Sciences, Beijing 100049, China}

\author[0000-0003-0274-3396]{Li-Ming Song}
\affil{State Key Laboratory of Particle Astrophysics, Institute of High Energy Physics, Chinese Academy of Sciences, Beijing 100049, China}
\affil{University of Chinese Academy of Sciences, Chinese Academy of Sciences, Beijing 100049, China}

\author{Wen-Jun Tan}
\affil{State Key Laboratory of Particle Astrophysics, Institute of High Energy Physics, Chinese Academy of Sciences, Beijing 100049, China}
\affil{University of Chinese Academy of Sciences, Chinese Academy of Sciences, Beijing 100049, China}

\author[0000-0003-0466-2223]{Ping Wang}
\affil{State Key Laboratory of Particle Astrophysics, Institute of High Energy Physics, Chinese Academy of Sciences, Beijing 100049, China}

\author[0000-0001-8664-5085]{Wang-Chen Xue}
\affil{State Key Laboratory of Particle Astrophysics, Institute of High Energy Physics, Chinese Academy of Sciences, Beijing 100049, China}
\affil{University of Chinese Academy of Sciences, Chinese Academy of Sciences, Beijing 100049, China}

\author[0000-0001-7599-0174]{Shu-Xu Yi}
\affil{State Key Laboratory of Particle Astrophysics, Institute of High Energy Physics, Chinese Academy of Sciences, Beijing 100049, China}

\author[0000-0001-9217-7070]{Yun-Wei Yu}
\affil{Institute of Astrophysics, Central China Normal University, Wuhan 430079, China}

\author[0009-0002-6411-8422]{Zheng-Hang Yu}
\affil{State Key Laboratory of Particle Astrophysics, Institute of High Energy Physics, Chinese Academy of Sciences, Beijing 100049, China}
\affil{University of Chinese Academy of Sciences, Chinese Academy of Sciences, Beijing 100049, China}

\author{Jin-Peng Zhang}
\affil{State Key Laboratory of Particle Astrophysics, Institute of High Energy Physics, Chinese Academy of Sciences, Beijing 100049, China}
\affil{University of Chinese Academy of Sciences, Chinese Academy of Sciences, Beijing 100049, China}

\author[0000-0002-8097-3616]{Peng Zhang}
\affil{State Key Laboratory of Particle Astrophysics, Institute of High Energy Physics, Chinese Academy of Sciences, Beijing 100049, China}
\affil{College of Electronic and Information Engineering, Tongji University, Shanghai 201804, China}

\author[0000-0001-5586-1017]{Shuang-Nan Zhang}
\affil{State Key Laboratory of Particle Astrophysics, Institute of High Energy Physics, Chinese Academy of Sciences, Beijing 100049, China}
\affil{University of Chinese Academy of Sciences, Chinese Academy of Sciences, Beijing 100049, China}

\author[0009-0008-6247-0645]{Wen-Long Zhang}
\affil{State Key Laboratory of Particle Astrophysics, Institute of High Energy Physics, Chinese Academy of Sciences, Beijing 100049, China}
\affil{School of Physics and Physical Engineering, Qufu Normal University, Qufu 273165, China}

\author[0000-0003-4673-773X]{Zhen Zhang}
\affil{State Key Laboratory of Particle Astrophysics, Institute of High Energy Physics, Chinese Academy of Sciences, Beijing 100049, China}

\author{Xiao-Yun Zhao}
\affil{State Key Laboratory of Particle Astrophysics, Institute of High Energy Physics, Chinese Academy of Sciences, Beijing 100049, China}

\author[0009-0001-7226-2355]{Chao Zheng}
\affil{State Key Laboratory of Particle Astrophysics, Institute of High Energy Physics, Chinese Academy of Sciences, Beijing 100049, China}
\affil{University of Chinese Academy of Sciences, Chinese Academy of Sciences, Beijing 100049, China}

\author{Shi-Jie Zheng}
\affil{State Key Laboratory of Particle Astrophysics, Institute of High Energy Physics, Chinese Academy of Sciences, Beijing 100049, China}

\begin{abstract}
Magnetar X-ray Burst (MXB) is usually composed of a single pulse or multiple pulses with rapid rise and brief duration mostly observed in hard X-ray (soft gamma-ray) band. 
Previous work studied the temporal behavior of some magnetar bursts and employed the Fast Rise Exponential Decay (FRED) model to fit pulses of MXB.
However, whether there is other kind of pulse shape has not been explored. 
In this study, we systematically examined light curve of MXBs from SGR J1935+2154 
detected by GECAM between 2021 and 2022. We find that there are different light curve morphologies. Especially, we discover a peculiar and new pattern, Exponential Rise and Cut-Off Decay (ERCOD), which is significantly different from FRED 
and could be well described by a mathematical function we proposed.
We find that MXBs with ERCOD shape are generally longer in duration, brighter in the peak flux\textcolor{black}{, and harder in spectrum than those with FRED.}
We note that the ERCOD shape is not unique to SGR J1935+2154 but also present in other magnetars. 
This new light curve pattern \textcolor{black}{may imply} a special burst and radiation mechanism of magnetar.
\end{abstract}


\section{Introduction} \label{sec:intro}
Magnetar, including soft gamma repeaters (SGRs) and anomalous X-ray pulsars (AXPs), is a special kind of neutron star 
with extremely strong magnetic field up to $10^{14}$ to $10^{15}$ G \citep{Duncan_1992ApJ,Kouveliotou_SGR}.
Such an extreme magnetic field can power the burst activity of magnetar \citep{Thompson_1993AIPC,Thompson_1995MNRAS}.
Indeed, SGRs are named for frequent burst events, emitting primarily in hard X-ray (or soft $\gamma$-ray) band \textcolor{black}{\citep{Woods_2006}}. These bursts are called X-ray bursts (XRBs) in some literatures. However, the acronym XRB has already been used to refer to the X-ray binary. To avoid such confusion, we suggest to name these bursts as Magnetar X-ray Burst (MXB), following the name convention of Magnetar Giant Flare (MGF).

The duration of MXB typically ranges from milliseconds to seconds \textcolor{black}{ \citep[e.g.][]{KONUS_CATALOG_2001,Woods_Thompson_2004,Lin_1935_2020}}. The common classification scheme divides MXB into short burst, intermediate flare, giant flare
\citep[e.g.][]{Woods_2006}. 
Recently the concept of super flare \citep{Zhang_2022} has been also proposed to account for peculiar precursor burst with QPO feature and higher energy release than giant flare \citep{Xiao_precursor_2022}.

In addition, compared to other pulsars, magnetars exhibit longer spin periods ($P$) , ranging approximately from 0.3 to 12 seconds, and faster spin-down rate ($\dot{P}$),
roughly between \(10^{-15}\) and \(10^{-10}\) seconds per second \citep[e.g.][]{Duncan_1992ApJ,Thompson_1996ApJ}.  

So far, the number of known magnetars has exceeded thirty \citep{McGill_Magnetar_Catalog_2014}, and the spatial distribution of magnetars in the Galaxy is strictly confined to the Galactic plane \citep{Kaspi_Beloborodov_2017}. There are also several magnetar bursts (giant flare) detected in nearby galaxies and were misidentified as gamma-ray burst (GRB), such as GRB 200415A in NGC 253 \citep{2020ApJ...899..106Y, 2021Natur.589..211S} and GRB 231115A in M82 \citep{2024Natur.629...58M}.

SGR J1935+2154 is the first (and unique as of this writing) magnetar firmly associated with fast radio bursts (FRBs) \citep[e.g.][]{CHIME_200428,2020ApJ...898L..29M,START2_200428,HXMT_200428,KW_0428,2021NatAs...5..401T,FRBaXRB_Lin_2020}. This association has sparked significant interest in exploring the underlying physical processes that link magnetar activity to FRBs. 
SGR J1935+2154 has a period $P$ of 3.25\,s and $\dot{P}$ of 3.45$\times10^{-11}$\,s$\cdot$s$^{-1}$\citep{1935_Spin_Evolution}. It has experienced several active episodes between 2014 and 2022 \citep[e.g.][]{SGR1935_2017,SGR1935_2020ApJ,SGR1935_Cai_2022,Xie_2022,Xie_1935_Catalog}.
Due to its prolific activity, association with FRB, and special spectral characteristics, SGR J1935+2154 has become a highly interesting object in the study of magnetar and FRB. Conducting in-depth research on it helps reveal the properties of magnetars and understand high-energy astrophysical phenomena \citep[e.g.][]{NS_model_15,NS_model_16,NS_model_17,NS_model_19,NS_model_20,NS_model_24,NS_model_25,NS_model_26,NS_model_27,FRBaXRB_2023,2023ApJ...950..121R}.

Light curve patterns often provide valuable information of the characteristics of the emission mechanism and central object being studied. \textcolor{black}{For example, in the case of transient phenomena like GRBs, light curves may offer a chronological record of central engine activity
\citep{Zhang_Zhang_Castro-Tirado_2016}.} The Fast Rise Exponential Decay (FRED) \citep{Norris_2005,Norris_1996} profile is a characteristic pattern observed in GRBs, which can describe either a single pulse or the entire burst \citep{2025ApJ...985..239Y}, depending on the analysis time period and the complexity of the burst. 
Studies suggest that the FRED profile in GRBs may arise from internal shocks in a relativistic wind: The rapid rise results from shock formation and synchrotron emission, while the exponential decay reflects electron cooling and decreasing radiative efficiency \citep{GRB_internal_shocks}. 
\textcolor{black}{In addition, the FRED profiles could also be explained by other scenarios, such as the external shocks \citep[e.g.][]{Fenimore_1996,Fenimore_1999}.}

Therefore, it is important to study the light curve patterns of MXB for the underlying physical processes of magnetar activity. Indeed, MXBs are widely thought to be produced by magnetic crustal breaks or magnetic reconnection, and their light curves should reflect the evolution of energy release process in magnetar.


Previous studies on the MXB light curve primarily focused on analyzing the full burst, rather than individual pulse within the burst \citep[e.g.][]{SGR1900_1999ApJ,SGR1806_2000ApJ}. Here we pay more attention to the pulse in the burst, just like the case in GRB study. We noticed that some pulses of MXBs of SGR J1935+2154 could be described by the FRED function, which is believed to be a basic light curve pattern found in GRB. Indeed, some research have scrutinized the temporal behavior of MXBs and have used the FRED model to fit pulses of MXB of \textcolor{black}{SGR J0501+4516 \citep{xiao_fred_fitting_2024} and SGR J1935+2154 \citep{Yang_Xiao_2025}}. However, it is unclear if there is other pattern than FRED in the light curve of MXB.
Indeed, we noticed that there are some MXB pulses that cannot be adequately described by FRED function, showing a systemic structure in the residual of the fit with FRED. 
This inspires us to investigate whether the MXB has different light curve patterns and whether the new light curve pattern can reveal new physics.

In this study, we examined light curves of 159 MXBs from SGR J1935+2154 \citep{Xie_1935_Catalog}, which were detected by the Gravitational Wave High-energy Electromagnetic Counterpart All-sky Monitor (GECAM) between 2021 and 2022. We systematically studied the morphologies of their light curves, and found that a subset of MXBs significantly deviate from the FRED pulse profile. 
To describe this new light curve pattern, we propose the Exponential Rise and Cut-Off Decay (ERCOD\footnote{We suggest to pronounce as \textipa{/"3:kOd/}.}) function, which is composed of two distinct stages: an exponential rise followed by a cut-off decay.
This unique light curve pattern is distinct from known pulse profile models such as the FRED.
Thus, in-depth exploration of these light curve patterns 
holds significant potential in unraveling the physical mechanisms governing the magentar burst.

This paper is structured as follows: In Section \ref{section2}, we describe the data reduction and the methods of fitting. \textcolor{black}{Section \ref{section3} shows the fitting, temporal and spectral analysis of FRED and ERCOD patterns of MXBs.} Finally, Section \ref{section4} and \ref{section5} present the discussion and summary.

\section{Observation and Data Reduction}\label{section2}

As of this writing, GECAM constellation comprises four all-sky space telescopes designed to detect X-ray and gamma-ray transients: GECAM-A and GECAM-B launched in December 2020 \citep{An_2022,2022RDTM....6...12L,gecam_b_li}, GECAM-C (also known as SATech-01/HEBS) launched in July 2022 \citep{gecam_c_zheng,2024ApJS..270....3X,2023NIMPA105668586Z}, and GECAM-D (also known as DRO/GTM) launched in March 2024 \citep{GTM_Wang_2024,gtm_feng}. 

Each of GECAM-A and GECAM-B microsatellite is equipped with 25 Gamma-Ray Detectors (GRDs), while GECAM-C houses 12 GRDs. \textcolor{black}{Each GRD module consists of a LaBr3:Ce scintillator readout by a dedicated SiPM array \citep{An_2022}.} All GRDs of GECAM-A/B and 10 out of 12 GRDs of GECAM-C \citep{gecam_c_zheng,2024ApJS..270....3X,2023NIMPA105668586Z} are configured with two readout channels: a high gain (HG) channel and a low gain (LG) channel. 
The HG channel, designed for the detection of low-energy photons, covers an energy range of approximately 8 keV to 300 keV (between 2021 and 2022) for GECAM-B and 6 keV to 300 keV for GECAM-C. The LG channel, designed for the detection of high-energy photons, covers an energy range of approximately 50 keV to 6000 keV for GECAM-B and 160 keV to 6000 keV for GECAM-C.
GECAM-D is equipped with 5 Gamma-ray Transient Probes (GTPs) that utilize a NaI(Tl) scintillator coupled with a SiPM array \citep{gtm_feng}. All GTPs have a single readout channel working in an energy range of about 10 keV to 1 MeV.

\textcolor{black}{Typically, MXBs emit photons in the X-ray to soft gamma-ray energy bands. Given that the HG channel of GECAM-B GRD detectors covers the hard X-ray to gamma-ray band from 8 to 300 keV,} the light curve analysis was performed using the HG channel of 25 GRD. \textcolor{black}{
Because the cumulative irradiation damage sustained during in-orbit operations, the gain of the SiPMs has gradually decreased and the dark noise of SiPM increased, 
leading to a gradual increase in the lower bound of the detection energy range \citep{2024RAA....24j4005Q}. To make best use of the energy range for spectral fit, we adopted specific minimum energy bound for each MXB 
while using a same value of upper limit of 300 keV. To enhance counting statistics, we selected all GRDs with incident angle less than $ 60^{\circ}$. The energy ranges and detector selection adopted for spectral analysis are listed in Appendix, Table \ref{tab:t90_energy_grd}. }

Thanks to the highest time resolution (0.1 $\mu$s) among all GRB detectors \citep{Xiao_time_calibrations_2022}, GECAM GRDs are suited for studying MXBs with short durations. 
The duration of these GECAM MXBs typically ranges from a few milliseconds to a few seconds. Here, the time bin of MXB light curve is set to 5 ms by default.  

In this work, \textcolor{black}{the MXB samples of this study derived from the GECAM-B burst catalog of SGR J1935+2154 during the burst active episode from 2021 to 2022 \citep{Xie_1935_Catalog}, which consists of 159 MXBs. And the "Burst ID" in the following figures and tables are consistent with SGR J1935+2154 burst list of GECAM-B.}
Among all MXBs detected by GECAM, we selected 23 representative MXB samples (shown in Figure \ref{fig:MXB_RECD}) and implemented extensive analysis. 
\textcolor{black}{For these MXBs, light-curve extraction, spectrum reduction, and response-matrix generation from GECAM data are all performed with \textit{GECAMTools-v20240531} \footnote{\url{https://github.com/zhangpeng-sci/GECAMTools-Public/}}.}
The light curves of 23 MXBs are fitted with the FRED pattern or ERCOD pattern (equations \ref{eq:fred} and \ref{eq:recd} respectively).
We employed the maximum likelihood method, i.e. maximizing the $ \ln\mathcal{L} = \sum_{i}[\text{DATA}_i \ln(\text{MODEL}_i) - \text{MODEL}_i]$
by assuming a Poisson distribution of the observed counts. The posteriors of the fitted parameters are derived with a Monte Carlo Markov Chain (MCMC) method. The fitted parameters are listed in extended data Table \ref{tab:fred_fit_par} and Table \ref{tab:ercod_par_fit}.

\section{light curve patterns and data analysis} \label{section3}
We systematically examined the light curves of 159 MXBs of SGR J1935+2154 detected by GECAM-B from 2021 to 2022, and identified that there are a variety of patterns in their light curves: some pulses of MXB could be fit with FRED function, which is the common pulse shape usually observed in GRB light curves. More interestingly, we find that some MXB light curves cannot be described by FRED. This type of light curve pattern features a cut-off decay as opposed to the exponential decay with slower and slower decay rates in the FRED function. 
Namely, it can be described by an exponential rising phase and sudden cut-off decay, i.e. ERCOD function. In addition, we note that many MXBs cannot be described by a single FRED or ERCOD function, but \textcolor{black}{possibly by} a combination of them. \textcolor{black}{We note that, the primary objective of this work is to investigate the general shape of the light curves of the pulse, rather than the fine and small structures in the light curve.}

\subsection{FRED light curve pattern}

The FRED shape is proposed to describe the typical light curve of GRB \citep{Norris_2005,Norris_1996}. 
The FRED pulse profile is known to be described in many forms \citep{Norris_2005, Norris_1996}, and we focus on the form consisting of the exponential function \citep{Norris_1996}.
\begin{equation}
\label{eq:fred}
I(t) = 
\begin{cases} 
A \cdot e^{-\left(\frac{|t-t_{max}|}{\sigma_1}\right)^\nu} & , t \leq t_{max} \\
A \cdot e^{-\left(\frac{|t-t_{max}|}{\sigma_2}\right)^\nu} & , t > t_{max}
\end{cases}
\end{equation}

FRED function has several key parameters: the maximum time point of the rising and decaying phases ($t_{\rm max}$), as well as the timescale parameters ($\sigma_1$ and $\sigma_2$) that measure the durations of the two phases, and the pulse shape modulation parameter ($\nu$), the normalization parameter (\textit{A}).

We fitted the pulses of some MXBs using the FRED function described by Equation \ref{eq:fred}.
The FRED shape can be used to describe the shape of a single pulse or the entire burst, depending on the time period of analysis and the complexity of the burst. In practical observations, some MXBs may exhibit more complex temporal structures, including multiple FRED-shape pulses overlapping or with some gaps. 
These patterns of MXB light curves are similar to GRB pulse profiles, except that the former usually has a shorter duration.

We note that, recently \cite{xiao_fred_fitting_2024} conducted a comprehensive temporal analysis of 27 MXBs detected by the \textit{Fermi} Gamma-ray Burst Monitor (\textit{Fermi}/GBM) from SGR J0501+4156, and also fitted these pulses with FRED function. They found that their results favor a magnetospheric origin, and that similarities with GRB imply that they share similar radiation mechanisms, such as magnetic reconnection processes.

\subsection{ERCOD light curve pattern}

Unlike the gradual decay of FRED shape, some MXBs exhibit a distinct behavior during their decay phase of the light curve. Specifically, these light curves exhibit an outwardly convex profile, characterized by a sharp cut-off decay rather than a gradual decay. In addition, this light curve pattern is not observed in some peculiar energy band but in all energy bands of GECAM.
Such light curve pattern even cannot be fitted with a single FRED. Fitting with multiple FRED pulses will cause fine tune parameter problem. This results indicate that it is a new basic light curve pattern.

To accurately model this peculiar light curve, we propose the ERCOD function, as shown in Equation\ref{eq:recd}. It is composed of an exponential rise phase and a cut-off decay phase: 
\begin{equation}
\label{eq:recd}
I(t) = 
\begin{cases} 
A \cdot e^{\left(-\frac{|t - t_{\max}|}{\sigma_1}\right)^\nu} & , t \leq t_{\max}, \\
A \cdot B \cdot\left[1 - e^{\left(-\frac{|t - t_{\text{cut}}|}{\sigma_2}\right)}\right] & , t_{\max} < t \leq t_{\text{cut}}
\end{cases}
\end{equation}
where $B = \left[1 - e^{\left(-\frac{|t_{\text{cut}} - t_{\max}|}{\sigma_2}\right)}\right]^{-1}$.
The difference between ERCOD and FRED lies in the presence of not only the maximum time ($t_{\max}$) but also a cut-off time ($t_{\rm cut}$), along with the time scale parameters ($\sigma_1$ and $\sigma_2$). In addition, the pulse shape modulation parameter of the rising stage ($\nu$) and the normalization parameter (\textit{A}) are also included. The parameter $B$ in the ERCOD model represents the normalization factor for the cut-off decay phase, ensuring continuity between the rise and decay stages.

We note that, although the mathematical formulation of this ERCOD model may have strong implication on the underlying physical processes to produce such pattern, the above formulation is not the only one that can describe the ERCOD pattern, just like the case for FRED pattern. Nevertheless, the sudden decay phase in the ERCOD pattern probably suggests a fundamentally different physical mechanism compared to the gradual decay of the FRED function.

\subsection{Temporal analysis}

The fitting parameters of the FRED pattern and the ERCOD pattern of the GECAM MXBs are shown in Table \ref{tab:fred_fit_par} and Table \ref{tab:ercod_par_fit}, respectively. The rise and decay time scales ($t_{\rm rise}$ and $t_{\rm decay}$), included in the above tables, represent the time for the pulse from the half maximum to maximum value ($t_{\rm rise}=t_{\rm max}-t_{\rm rise,1/2}$ and $t_{\rm decay}=t_{\rm decay,1/2}-t_{\rm max}$, where $t_{\rm rise,1/2}$ and $t_{\rm decay,1/2}$ are the times when the signal reaches half of the maximum value in the rise and decay stages). 
The errors of $t_{\rm rise}$ and $t_{\rm decay}$ were calculated by simulating multiple sets of light curves using the fitting parameters generated through the MCMC fitting process.

\begin{figure*}[!htbp] 
\centering
\begin{tabular}{ccc}    
    \begin{overpic}[width=\textwidth]{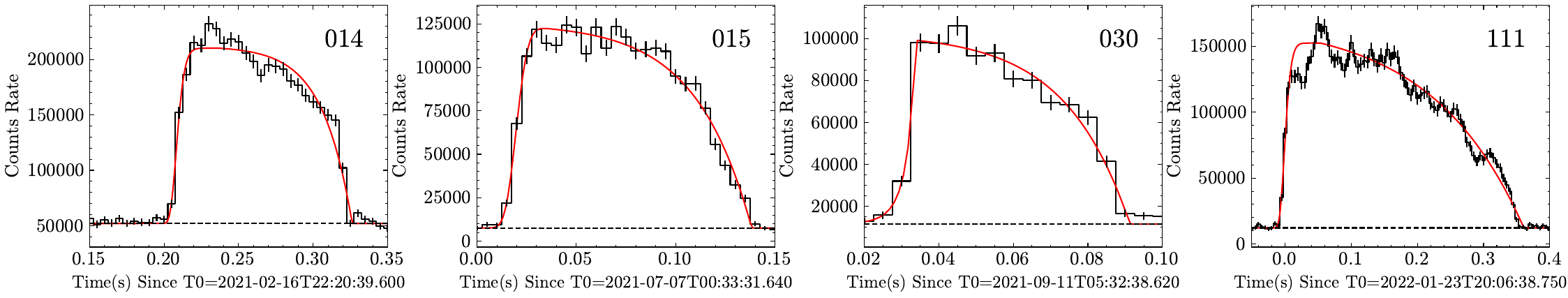}
    \put(-3, 18){\bf a}
    \end{overpic}\\
    \begin{overpic}[width=0.49\textwidth]{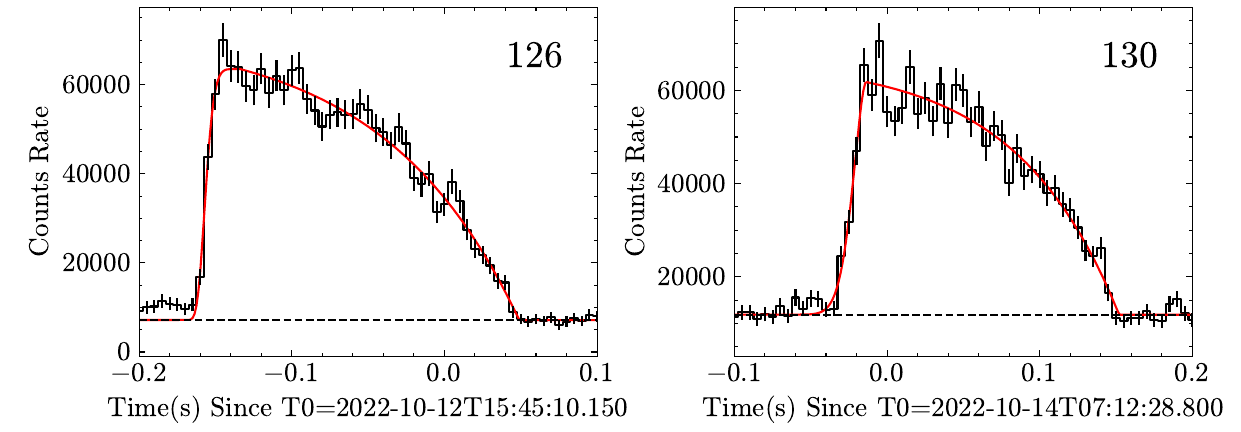}
    \end{overpic}\\
    
    \begin{overpic}[width=\textwidth]{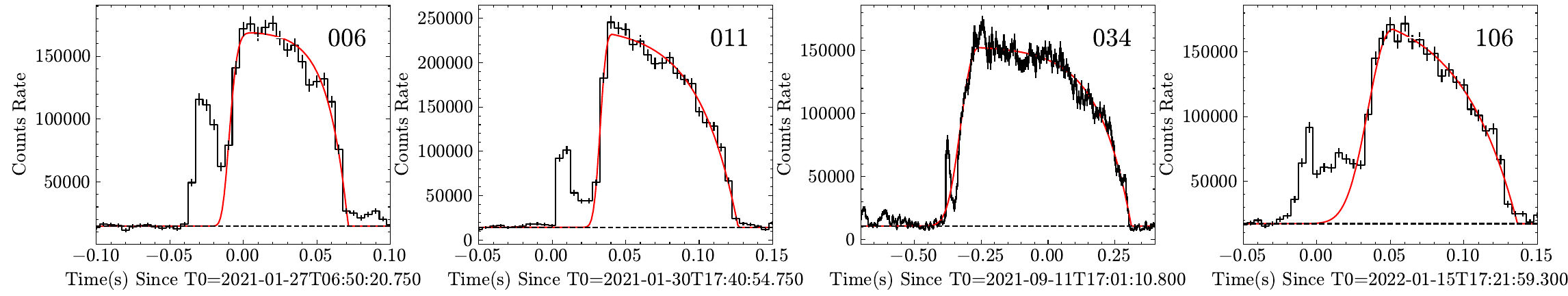}
    \put(-3, 18){\bf b}
    \end{overpic}\\
    
   \begin{overpic}[width=\textwidth]{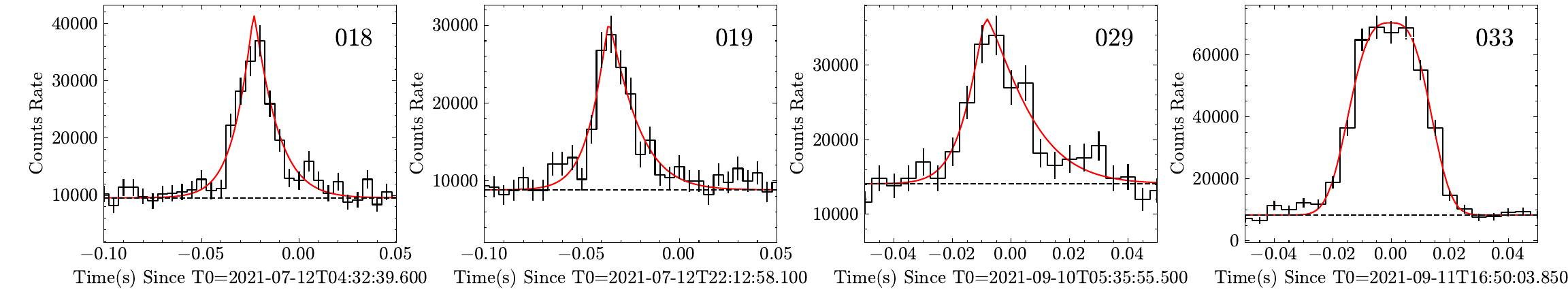}
   \put(-3, 18){\bf c}
   \end{overpic}\\

\end{tabular}
\caption{\noindent\textbf{Light curves of some MXBs from SGR J1935+2154 detected by GECAM.} The red lines are the fitting results of the ERCOD or FRED model based on the MCMC method.
\textbf{a}: Typical MXBs with ERCOD light curve pattern.
\textbf{b}: MXBs with ERCOD light curve pattern and an accompanying initial pulse. The initial pulse is ignored in the fitting.
\textbf{c}: MXBs with FERD light curve pattern. 
}
\label{fig:MXB_RECD}
\end{figure*}

Among of the 159 MXBs, we find there are 10 MXBs that could be well fit with ERCOD function (see Figure \ref{fig:MXB_RECD}) and 6 MXBs that are potentially fit with ERCOD (thus called ERCOD candidates). Thus, approximately $10\%$ of the GECAM-detected MXBs consist of ERCOD shape light curves.
ERCOD light curves from typical to less typical samples are shown in Figure \ref{fig:MXB_RECD}(\textbf{a}) and Figure \ref{fig:MXB_RECD}(\textbf{b}), respectively. 
Figure \ref{fig:MXB_RECD}(\textbf{a}) shows a very typical ERCOD without other pulse structure contamination. These light curves can be directly fitted with a basic pattern ERCOD. In contrast, Figure \ref{fig:MXB_RECD}(\textbf{b}) shows ERCOD pattern with an accompanying pulse. For these light curves, the initial pulse segment is ignored during the fitting process.

Although the catalog includes a significant number of MXBs that exhibit FRED patterns, only a portion of them is selected to be displayed in Figure \ref{fig:MXB_RECD}(\textbf{c}) . About 13 MXBs with typical FRED patterns in the catalog have been fitted. The samples of the FRED pattern and the fitting results are shown in Figure \ref{fig:MXB_RECD}(\textbf{c}) , where the red lines are the fitted results based on the MCMC method.

We stress that not all pulses of MXBs can be adequately fitted by the ERCOD or FRED functions. Some MXBs exhibit complicated light curve structures, indicating that a more sophisticated model or a combination of multiple models is necessary to achieve a satisfactory fit of the light curve.

\subsection{Spectral analysis}

\textcolor{black}{
Following previous studies of MXB spectrum \citep[e.g.][]{2023ApJ...950..121R,2024ApJ...969...38R}, we used Blackbody (BB) and Cutoff Powerlaw (CPL) to fit the time-integrated spectrum within the T90 interval of 23 MXB samples using spectral analysis tool \texttt{ELISA} \citep{elisa2024}. We find that BB provide statistically better fitting than CPL (with all parameters free). 
The main reason is that the low energy threshold of GECAM-B data of these bursts is not sufficiently below the $E_{p}$ of CPL, making the parameter $\alpha$ of CPL unconstrained. Since the $\alpha$ of CPL model usually varies around 1 and a CPL model with $\alpha=1$ will not significantly affect the spectral fit of GECAM-B spectrum of these MXBs, we simply fixed $\alpha$ of CPL to 1, denoted as CPL ($\alpha=1$) hereafter, to estimate the $E_{p}$ of CPL model, which is the parameter of interest in this work.} 

Spectral fitting results of BB and CPL ($\alpha=1$) are listed in Table \ref{tab:spec_fitting_result}. The selection of time, energy range and detectors for spectral analysis are compiled in Table \ref{tab:t90_energy_grd}. From these results, one can find that the majority of FRED samples are well described by both the BB and CPL ($\alpha=1$) model, which is evident from the fit statistics ($stat/dof \approx 1$, lower $WAIC$) and residual structure. 
Most ERCOD samples can also be adequately fit by these two models. However, some ERCOD samples (e.g. No.34, No.111, and No.126 in Table \ref{tab:spec_fitting_result}) exhibit significant residual in the fitting, which probably indicates the existence of multiple spectral components or spectral evolution due to the relatively long duration of ERCOD burst.

\textcolor{black}{We compare the spectral parameters of ERCOD and FERD bursts for both BB and CPL ($\alpha=1$) models, as shown in Figure \ref{png:kt_flux}. We find that ERCOD bursts exhibit, on average, higher $kT$ values in BB model and higher $E_{\rm p}$ in CPL ($\alpha=1$) model than FRED ones, which indicates that ERCOD bursts generally have a higher spectral hardness compared to those with FRED pattern.}

\section{Discussion}\label{section4}

Why and how to produce these different light curve patterns are interesting and important question, through which we may probe the physics of magnetar, especially through this peculiar ERCOD shape light curve found in this work. We note that FRB 200428-associated X-ray burst detected by HXMT \citep{HXMT_200428} can be fitted with multiple FRED function rather than the ERCOD function.

We examine the possibility that ERCOD bursts result from the superposition of multiple FRED events or other simpler structures. Producing an ERCOD light curve several times brighter than a typical FRED would require multiple FREDs to occur almost simultaneously, making it difficult to explain the longer duration of the ERCOD burst. Alternatively, to extend the duration, distributing multiple FREDs over a longer interval would likely smooth out the light curve, preventing the fast rise characteristic of ERCODs. It also requires a fine tune of the parameters of these FREDs. Therefore, we conclude that it is very unlikely that the ERCOD is a superposition of multiple FREDs or other elementary light curve pattern.

\begin{figure*}
\centering
\begin{tabular}{ccc}
    \begin{overpic}[width=0.49\textwidth]{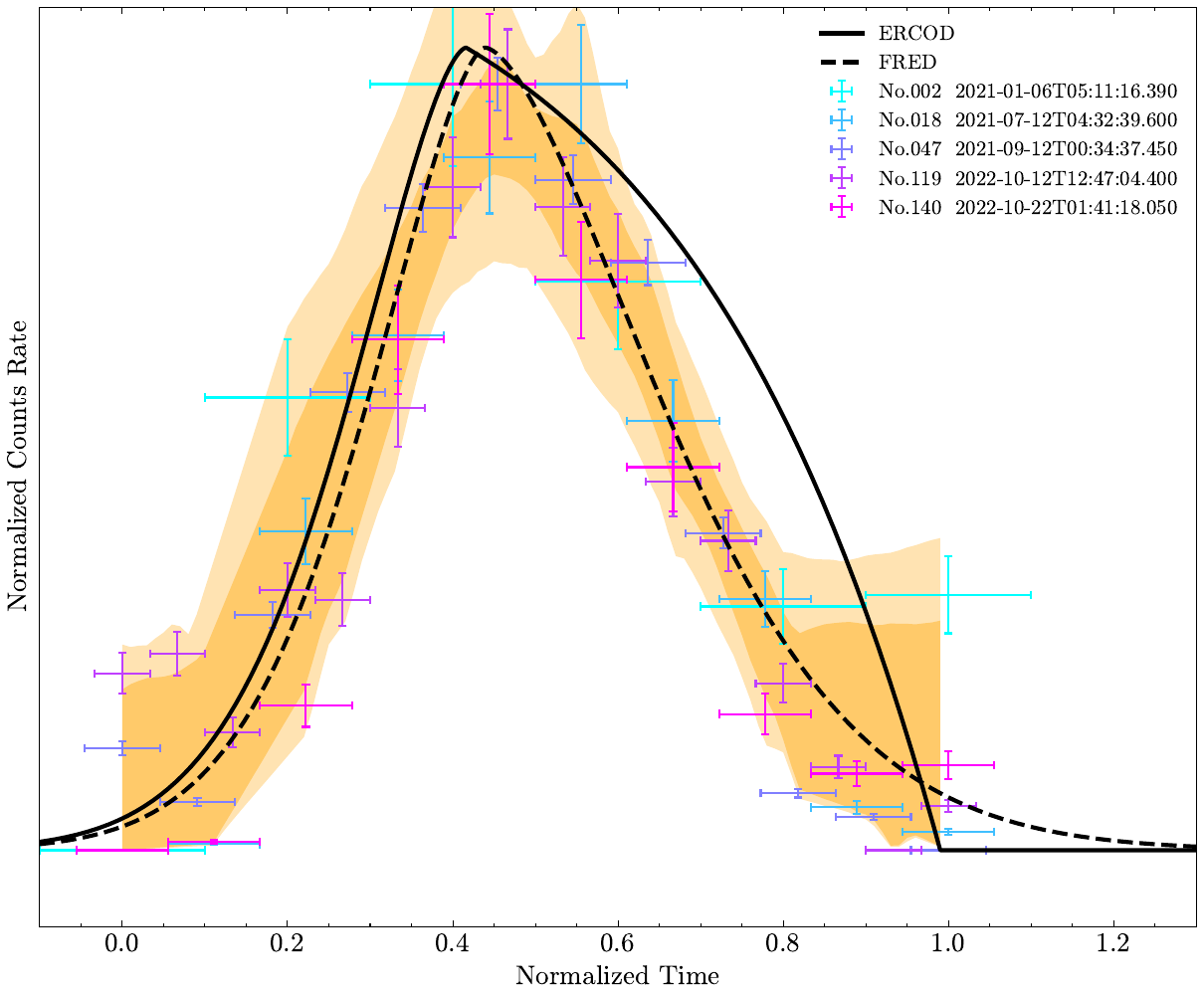}
        \put(-3, 80){\bf a}
    \end{overpic}
    &
    \begin{overpic}[width=0.49\textwidth]{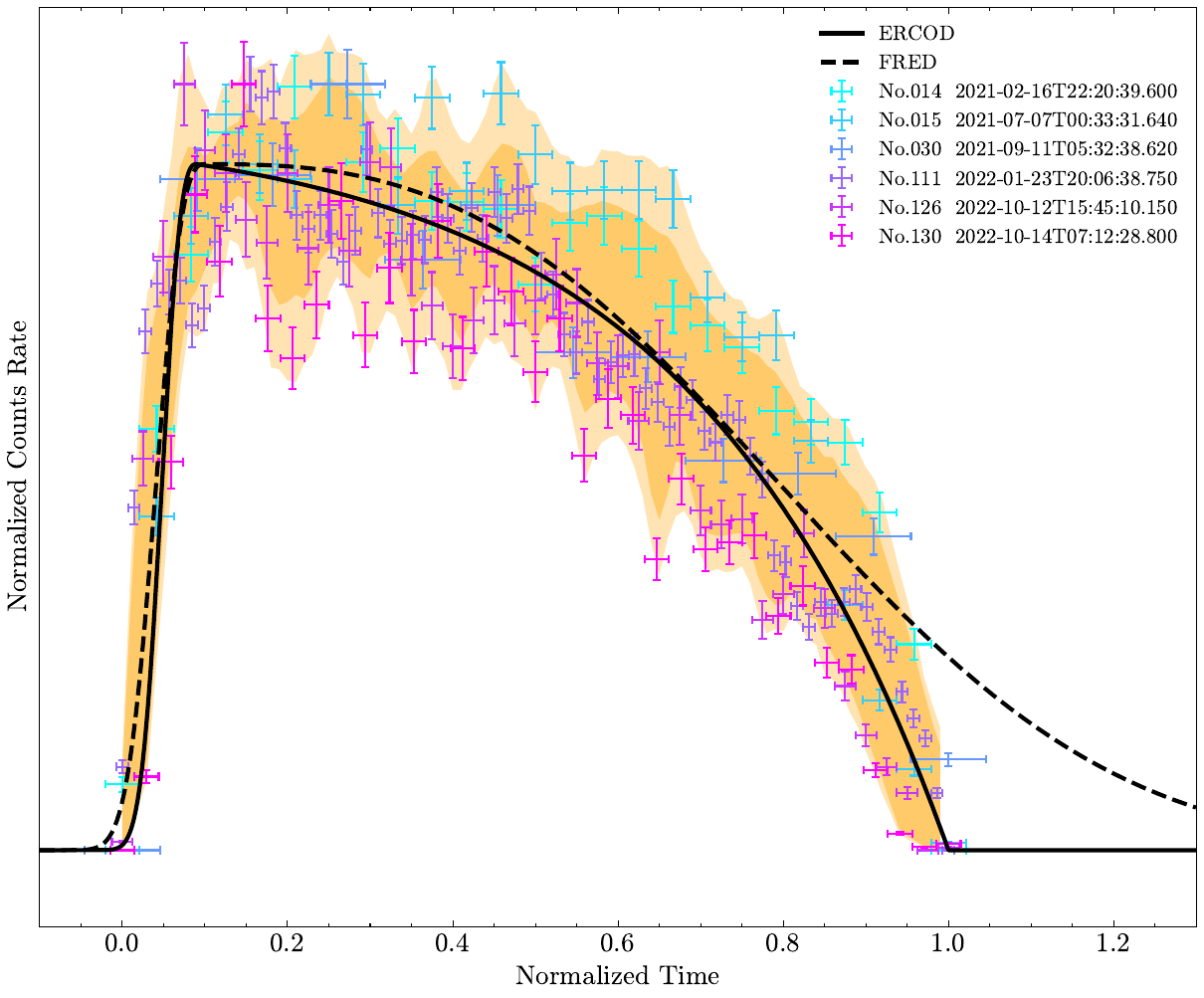}
        \put(-3, 80){\bf b}
    \end{overpic}
\end{tabular}
\caption{\noindent\textbf{ Fitting of normalized light curves.} 
The FRED (dashed line) and ERCOD (solid line) shape are plot over the normalized light curves.
\textbf{a}: Normalized light curves of MXBs with FRED light curve pattern.
\textbf{b}: Normalized light curves of MXBs with ERCOD light curve pattern.
}
\label{png:fitting_norm}
\end{figure*}

The normalized light curves of Figure \ref{png:fitting_norm} highlight the differences of the two patterns. It appears that the FRED model, represented by Equation \ref{eq:fred}, does not adequately describe the declining trend and the abrupt cut-off of the tail in ERCOD light curve pattern. Both the FRED function (dashed line) and the ERCOD function (solid line) are plotted. The shaded areas indicate the 1$\sigma$ and 3$\sigma$ ranges. In contrast, Equation \ref{eq:recd}, which depicts the ERCOD pattern, cannot fit the exponential decay of the FRED pattern adequately. Consequently, this clearly demonstrates the fundamental morphological distinctions between these two light curve patterns, indicating that the cut-off decay of ERCOD and the exponential decay of FRED represent two entirely distinct fundamental light curve patterns.

\begin{figure*}
\centering
\begin{tabular}{c}
    \begin{overpic}[width=0.49\textwidth]{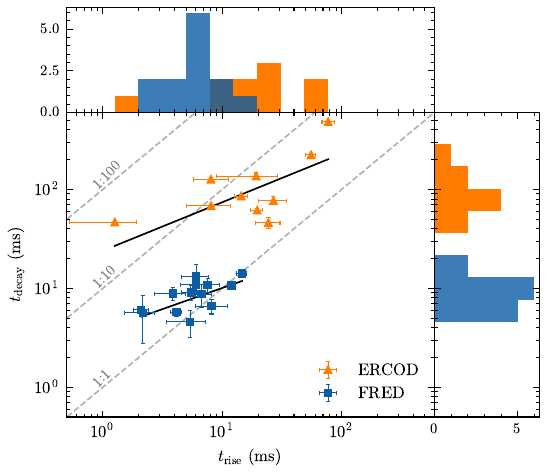}
    \end{overpic}
\end{tabular}
\caption{\noindent\textbf{The distribution of $t_{\rm rise}$ and $t_{\rm decay}$ for MXBs with ERCOD (orange triangles) and FRED (blue squares) patterns.} The black lines are the fitting results to the data separately. The gray dotted lines indicate the ratio between $t_{\rm rise}$ and $t_{\rm decay}$.
}
\label{png:tr_td}
\end{figure*}

\begin{figure*}
\centering
\begin{tabular}{cc}
    \begin{overpic}[width=0.49\textwidth]{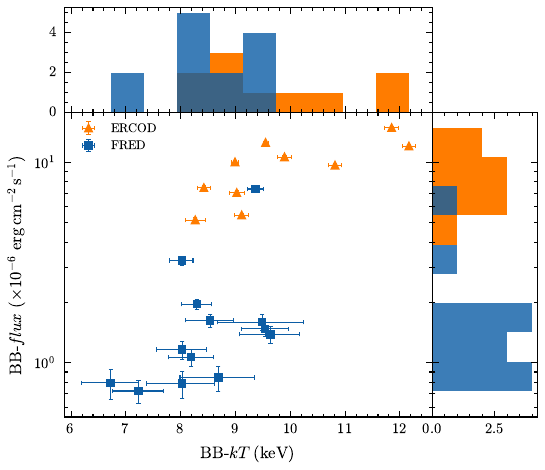}
        \put(-2, 85){\bf a}
    \end{overpic}
    &
    \begin{overpic}[width=0.49\textwidth]{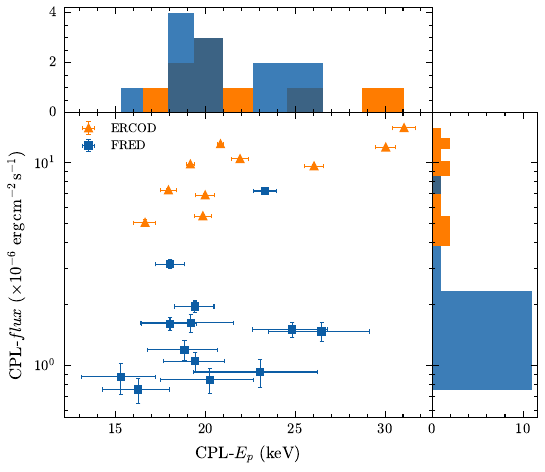}
        \put(-2, 85){\bf b}
    \end{overpic}
\end{tabular}
\caption{\noindent\textbf{Spectral properties of ERCOD (orange triangles) and FRED (blue squares) samples.} 
\textbf{a}: Distribution of blackbody temperature ($kT$) and flux (obtained from the BB model). 
\textbf{b}: Distribution of $E_{\rm p}$ and flux (obtained from the CPL ($\alpha=1$) model).}
\label{png:kt_flux}
\end{figure*}

The two patterns can clearly be divided into two distinct clusters based on temporal properties. The rise and decay timescales ($t_{\rm rise}$ and $t_{\rm decay}$) of the FRED and ERCOD samples are shown in Figure \ref{png:tr_td} (\textbf{a})) and Tables \ref{tab:fred_fit_par} and \ref{tab:ercod_par_fit}. The distinction of the ERCOD and FRED bursts is clearly seen in these results, indicating that there are fundamental difference between these two types of MXBs. In general, the ERCOD light curve pattern exhibits a significantly much longer decay timescale compared to its rise timescale. Moreover, the $t_{\rm decay}/t_{\rm rise}$ of the ERCOD is also somewhat larger than that of the FRED bursts. 

\begin{figure*}
\centering
\begin{tabular}{cc}
    \begin{overpic}[width=0.49\textwidth]{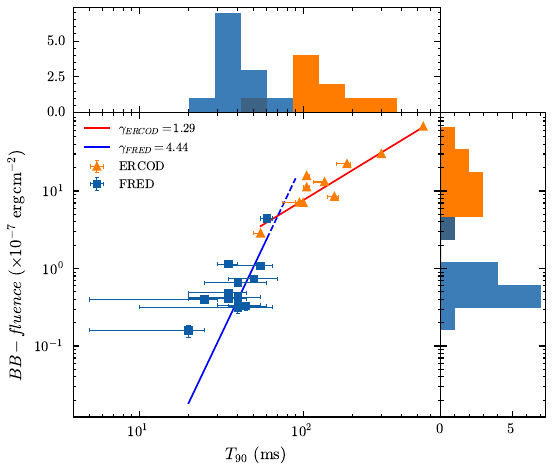}
        \put(-2, 85){\bf a}
    \end{overpic}
    &
    \begin{overpic}[width=0.49\textwidth]{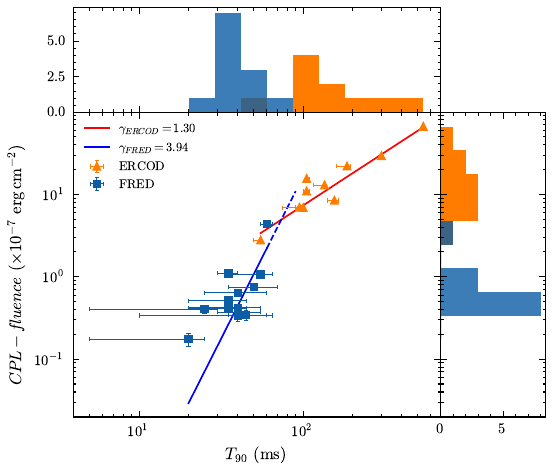}
        \put(-2, 85){\bf b}
    \end{overpic}
\end{tabular}
\caption{\noindent\textbf{Distribution of fluence and duration ($T_{90}$)for MXBs with ERCOD (orange triangles) and FRED (blue squares) patterns.}
\textbf{a}: Fluence derived from the BB model.
\textbf{b}: Fluence derived from the CPL ($\alpha=1$) model. 
The red and blue lines represent the fitting results for the ERCOD and FRED samples, respectively.}
\label{png:duration_fluence}
\end{figure*}

\textcolor{black}{More difference of these two type of MXBs can be seen from} their spectral properties (Figure \ref{png:kt_flux}), fluence and burst duration ($T_{90}$). \textcolor{black}{As mentioned above, we cannot determine whether BB or CPL is the intrinsic spectrum model for these MXBs based on the GECAM-B data in this wok. Thus we fit their spectrum with both models, and calculate their fluence based on spectral results with BB (Figure \ref{png:duration_fluence}(\textbf{a})) and CPL ($\alpha=1$) (Figure \ref{png:duration_fluence}(\textbf{b})), respectively. Since the calculation for fluence is for energy range of 35–300 keV, the fluences derived from both BB and CPL ($\alpha=1$) are well compatible to each other. 
It is evident that MXBs with the ERCOD light curve pattern have higher fluence and longer duration compared to those with the FRED pattern. Moreover, we fit the trend of fluence and duration with the power-law function ($fluence=A\cdot T_{90}^{\gamma}$) for the ERCOD and FRED samples respectively, as shown in Figure 5. There seems to be a systematic difference in $\gamma$ for ERCOD and FRED burst samples, which potentially implies that they have different energy-release rate.}


\section{Summary}\label{section5}
In this study, we systematically studied the light curve patterns of 159 MXBs from SGR J1935+2154 detected by GECAM-B from 2021 to 2022, and found that there are different patterns in the light curve. 
Especially, we discovered a very special light curve pattern for which we propose the ERCOD function to describe. As this light curve pattern cannot be reasonably fit with a combination of multiple basic functions (like FRED), we conclude that the ERCOD shape is a fundamental pattern in the MXB light curve, just like FRED.
As far as we know, this is the first time that such kind of light curve shape (i.e. exponentially rise followed by a cut-off decay) is reported in magnetar burst and other transients.

\textcolor{black}{
We emphasize that both FRED and ERCOD should be considered as fundamental pattern of MXB light curves, while it is possible that there are other fundamental patterns. The light curves of MXBs can be composed of one or more single or mixed these fundamental patterns. }
Only a small portion (about 10\%) of GECAM-detected MXBs could be well described by the ERCOD shape.

Moreover, it should be noted that this ERCOD type of light curve pattern is not unique to the MXBs of SGR J1935+2154; it is also evident in other magnetar bursts, such as the burst from SGR 1806-20 (see Figure 1 in \cite{1987ApJ...320L.105A}). Therefore, the ERCOD shape light curve pattern seems to be a common feature of magnetar bursts. 

\textcolor{black}{
We discussed the distinctions between the ERCOD and FRED light curve patterns in terms of temporal and spectral properties, and we find the the following conclusions: 
\begin{itemize}
 \item{The ERCOD pattern should be regarded as a new fundamental pattern, distinct to the FRED pattern.}
 \item{The ERCOD-shaped MXBs on average have higher luminosity, longer duration, larger decay timescale and harder spectrum in comparison to the FRED-shaped bursts.}
  \item{
  \textcolor{black}{With the current GECAM-B observation data, both the BB and CPL model can adequately fit the spectrum of most MXBs. 
  Joint observation of GECAM and soft X-ray telescopes is required to pin down the best spectral model.}}
\end{itemize}
} 

Finally, we note that the discovery of the peculiar ERCOD pattern of MXB light curve provides a new window to probe the physics of magnetar. 
Based on the observation results of this work, we envisage that, compared to the FRED bursts, the ERCOD pattern bursts probably caused by more powerful energy release process of magnetar.
However, the exact and detailed physics leading to the ERCOD light curve pattern, as well as the FRED pattern, is a very interesting question, which requires in-depth study.

\section*{Acknowledgments}
We appreciate the anonymous reviewers for valuable suggestions and comments. This work is supported by the National Natural Science Foundation of China (Grant No. 12273042, 12494572),
the Strategic Priority Research Program, the Chinese Academy of Sciences (Grant No. 
XDA30050000, 
XDB0550300 
),
the National Key R\&D Program of China (2021YFA0718500).
The GECAM (Huairou-1) mission is supported by the Strategic Priority Research Program on Space Science (Grant No. XDA15360000) of Chinese Academy of Sciences.


\begin{table*}[htbp]
\caption{\centering{Fitting parameters of FRED pattern of MXBs from SGR J1935+2154 detected by GECAM-B.}}
\tiny
\begin{tabular*}{\hsize}{@{}@{\extracolsep{\fill}}cccccccccc@{}}
\toprule
Burst ID	&	$\rm T_0$ (UTC)	&	$\rm \sigma_1~(\rm ms)$	&	$\rm \sigma_2~(\rm ms)$	&	$t_{\rm max}~(\rm ms)$  &	$\rm \nu$  &	$A$	&	$\rm bkg ~(\rm counts)$ &	$t_{\rm rise}~(\rm ms)$   & $t_{\rm decay}~(\rm ms)$   \\

\hline
018	&	 2021-07-12T04:32:39.600	&	 $10.0_{-2.0}^{+2.0}$	&	 $12.8_{-3.5}^{+3.3}$	&	 $-23.4_{-1.4}^{+1.6}$	&	 $1.15_{-0.26}^{+0.27}$	&	 $158_{-24}^{+25}$	&	 $47.80_{-0.35}^{+0.35}$	&	 $6.8_{-1.6}^{+1.6}$	&	 $9.2_{-3.0}^{+3.5}$ \\
019	&	 2021-07-12T22:12:58.100	&	 $10.2_{-2.4}^{+2.5}$	&	 $15.6_{-2.4}^{+2.4}$	&	 $-36.2_{-1.7}^{+1.6}$	&	 $1.15_{-0.12}^{+0.12}$	&	 $108_{-11}^{+11}$	&	 $44.22_{-0.33}^{+0.34}$	&	 $7.8_{-2.2}^{+2.5}$	&	 $12.5_{-1.8}^{+2.0}$ \\
026	&	 2021-09-10T03:24:47.150	&	 $3.1_{-0.4}^{+0.4}$	&	 $8.3_{-0.6}^{+0.5}$	&	 $2.11_{-0.30}^{+0.31}$	&	 $1.140_{-0.033}^{+0.033}$	&	 $590_{-40}^{+40}$	&	 $56.7_{-0.4}^{+0.4}$	&	 $2.25_{-0.26}^{+0.34}$	&	 $6.1_{-0.5}^{+0.5}$ \\
029	&	 2021-09-10T05:35:55.500	&	 $8.0_{-2.4}^{+2.4}$	&	 $17_{-4}^{+4}$	&	 $-8.5_{-1.5}^{+1.7}$	&	 $1.27_{-0.32}^{+0.35}$	&	 $113_{-18}^{+20}$	&	 $70.5_{-0.4}^{+0.4}$	&	 $6.0_{-2.2}^{+1.7}$	&	 $12.6_{-3.5}^{+5.0}$ \\
033	&	 2021-09-11T16:50:03.850	&	 $16.7_{-1.4}^{+1.4}$	&	 $15.9_{-1.3}^{+1.3}$	&	 $-0.1_{-1.3}^{+1.3}$	&	 $3.11_{-0.33}^{+0.33}$	&	 $309_{-12}^{+11}$	&	 $41.65_{-0.33}^{+0.34}$	&	 $14.3_{-1.0}^{+0.8}$	&	 $14.5_{-1.0}^{+1.0}$ \\
047	&	 2021-09-12T00:34:37.450	&	 $16.0_{-0.7}^{+0.7}$	&	 $14.4_{-0.7}^{+0.7}$	&	 $-8.5_{-0.5}^{+0.5}$	&	 $1.32_{-0.07}^{+0.07}$	&	 $896_{-31}^{+30}$	&	 $66.2_{-0.4}^{+0.4}$	&	 $12.1_{-0.7}^{+0.7}$	&	 $10.8_{-0.7}^{+0.5}$ \\
067	&	 2021-09-29T23:41:12.245	&	 $4.9_{-1.5}^{+1.4}$	&	 $10.9_{-1.8}^{+1.8}$	&	 $3_{-1}^{+1}$	&	 $1.32_{-0.25}^{+0.26}$	&	 $149_{-21}^{+21}$	&	 $61.2_{-0.4}^{+0.4}$	&	 $3.8_{-1.4}^{+1.8}$	&	 $7.9_{-1.3}^{+1.2}$ \\
077	&	 2022-01-09T07:39:10.700	&	 $7.3_{-1.4}^{+1.4}$	&	 $12.4_{-1.9}^{+2.0}$	&	 $-2.8_{-0.9}^{+0.9}$	&	 $1.09_{-0.07}^{+0.07}$	&	 $165_{-17}^{+17}$	&	 $56.4_{-0.4}^{+0.4}$	&	 $4.9_{-1.1}^{+1.1}$	&	 $9.5_{-1.7}^{+1.6}$ \\
091	&	 2022-01-14T19:45:08.100	&	 $2.6_{-1.7}^{+2.0}$	&	 $7.0_{-3.5}^{+6.0}$	&	 $-25.4_{-1.1}^{+1.4}$	&	 $1.0_{-0.5}^{+0.9}$	&	 $174_{-35}^{+34}$	&	 $76.1_{-3.4}^{+14.0}$	&	 $2.0_{-1.0}^{+0.8}$	&	 $7.1_{-3.3}^{+3.2}$ \\
115	&	 2022-05-21T23:32:58.800	&	 $8.5_{-3.5}^{+4.0}$	&	 $7.7_{-3.2}^{+3.2}$	&	 $-11.1_{-1.0}^{+0.9}$	&	 $0.87_{-0.27}^{+0.26}$	&	 $189_{-21}^{+70}$	&	 $60.9_{-0.5}^{+0.5}$	&	 $5.1_{-2.4}^{+2.1}$	&	 $4.9_{-2.1}^{+2.1}$ \\
117	&	 2022-05-24T17:10:18.550	&	 $6.0_{-0.8}^{+0.8}$	&	 $7.9_{-1.3}^{+1.3}$	&	 $-17.9_{-0.5}^{+0.5}$	&	 $1.15_{-0.13}^{+0.12}$	&	 $256_{-30}^{+30}$	&	 $58.4_{-0.4}^{+0.4}$	&	 $4.3_{-0.9}^{+0.9}$	&	 $6.0_{-1.4}^{+1.2}$ \\
132	&	 2022-10-14T11:27:32.750	&	 $10.2_{-2.2}^{+2.3}$	&	 $7.7_{-1.8}^{+1.7}$	&	 $-9.3_{-1.7}^{+2.1}$	&	 $2.3_{-0.8}^{+0.8}$	&	 $164_{-16}^{+19}$	&	 $46.54_{-0.35}^{+0.35}$	&	 $8.5_{-1.3}^{+1.9}$	&	 $6.3_{-1.1}^{+1.1}$ \\
140	&	 2022-10-22T01:41:18.050	&	 $6.7_{-2.6}^{+2.6}$	&	 $12.8_{-2.5}^{+2.6}$	&	 $-35.9_{-2.1}^{+2.0}$	&	 $2.4_{-0.9}^{+0.8}$	&	 $84_{-12}^{+12}$	&	 $38.71_{-0.30}^{+0.30}$	&	 $5.4_{-1.4}^{+1.8}$	&	 $11.8_{-2.0}^{+2.0}$ \\

\hline
\end{tabular*}
\label{tab:fred_fit_par}
\end{table*}

\begin{table*}[htbp]
\caption{\centering{Fitting parameters of ERCOD pattern of MXBs from SGR J1935+2154 detected by GECAM-B.}}
\tiny
\begin{tabular*}{\hsize}
{@{}@{\extracolsep{\fill}}ccccccccccc@{}}
\hline
\toprule

Burst ID	&	$\rm T_0$ (UTC)	&	$\rm \sigma_1~ (ms)$	&	$\rm \sigma_2~ (ms)$	&	$t_{\rm max} ~(\rm ms)$  & $t_{\rm cut} ~(\rm ms)$  &	$\rm \nu$  &	$A$	& $\rm bkg (counts)$	& $t_{\rm rise}~(\rm ms)$   & $t_{\rm decay}~(\rm ms)$   \\

\hline
006	&	2021-01-27T06:50:20.750	&	 $18_{-8}^{+7}$	&	 $13.6_{-1.0}^{+1.0}$	&	 $7_{-8}^{+7}$	&	 $71.30_{-0.27}^{+0.28}$	&	 $4.7_{-2.7}^{+2.5}$	&	 $769_{-12}^{+12}$	&	 $73.5_{-0.5}^{+0.4}$	&	 $18_{-9}^{+10}$	&	 $53_{-9}^{+9}$ \\
011	&	2021-01-30T17:40:54.750	&	 $10_{-4}^{+4}$	&	 $27.3_{-1.7}^{+1.7}$	&	 $41_{-4}^{+4}$	&	 $126.63_{-0.33}^{+0.33}$	&	 $3.0_{-1.5}^{+1.5}$	&	 $1088_{-16}^{+17}$	&	 $71.9_{-0.4}^{+0.4}$	&	 $9_{-4}^{+5}$	&	 $67_{-5}^{+4}$ \\
014	&	2021-02-16T22:20:39.600	&	 $29_{-8}^{+8}$	&	 $19.9_{-1.7}^{+1.6}$	&	 $236_{-8}^{+8}$	&	 $326.3_{-0.5}^{+0.5}$	&	 $8.3_{-2.5}^{+2.5}$	&	 $792_{-12}^{+12}$	&	 $259.8_{-0.8}^{+0.8}$	&	 $26_{-8}^{+7}$	&	 $77_{-7}^{+8}$ \\
015	&	2021-07-07T00:33:31.640	&	 $17.0_{-3.5}^{+4.0}$	&	 $28.1_{-2.0}^{+2.0}$	&	 $35.3_{-3.4}^{+4.0}$	&	 $138.3_{-0.5}^{+0.6}$	&	 $4.0_{-0.9}^{+0.9}$	&	 $574_{-9}^{+9}$	&	 $38.02_{-0.31}^{+0.31}$	&	 $15.6_{-3.1}^{+2.9}$	&	 $84.5_{-2.5}^{+2.8}$ \\
030	&	2021-09-11T05:32:38.620	&	 $2.3_{-1.3}^{+1.4}$	&	 $17.3_{-2.2}^{+2.2}$	&	 $33.6_{-1.4}^{+1.6}$	&	 $91.3_{-0.5}^{+0.5}$	&	 $0.95_{-0.31}^{+0.50}$	&	 $439_{-13}^{+13}$	&	 $57.5_{-0.4}^{+0.4}$	&	 $1.6_{-1.4}^{+1.4}$	&	 $46.4_{-2.2}^{+2.7}$ \\
034	&	2021-09-11T17:01:10.800	&	 $89_{-10}^{+10}$	&	 $122.7_{-3.1}^{+3.0}$	&	 $-256_{-9}^{+9}$	&	 $311.6_{-0.9}^{+0.9}$	&	 $2.47_{-0.28}^{+0.28}$	&	 $708_{-4}^{+4}$	&	 $53.4_{-0.4}^{+0.5}$	&	 $76_{-10}^{+11}$	&	 $485_{-9}^{+9}$ \\
106	&	2022-01-15T17:21:59.300	&	 $22.9_{-1.7}^{+1.8}$	&	 $38.7_{-3.4}^{+3.4}$	&	 $51.9_{-1.6}^{+1.6}$	&	 $136.5_{-0.5}^{+0.5}$	&	 $2.18_{-0.15}^{+0.15}$	&	 $753_{-13}^{+12}$	&	 $85.5_{-0.5}^{+0.5}$	&	 $18.8_{-2.1}^{+1.5}$	&	 $62.9_{-2.0}^{+2.2}$ \\
111	&	 2022-01-23T20:06:38.750	&	 $55_{-7}^{+7}$	&	 $139_{-5}^{+5}$	&	 $54_{-7}^{+7}$	&	 $360.9_{-1.0}^{+1.0}$	&	 $9.0_{-1.3}^{+1.2}$	&	 $701.0_{-0.8}^{+0.8}$	&	 $61.1_{-0.4}^{+0.4}$	&	 $53_{-5}^{+6}$	&	 $226_{-5}^{+5}$ \\
126	&	2022-10-12T15:45:10.150	&	 $25_{-16}^{+13}$	&	 $89_{-9}^{+10}$	&	 $-132_{-15}^{+12}$	&	 $49.1_{-1.5}^{+1.6}$	&	 $7_{-5}^{+4}$	&	 $281_{-7}^{+7}$	&	 $35.83_{-0.32}^{+0.31}$	&	 $19_{-8}^{+12}$	&	 $135_{-13}^{+10}$ \\
130	&	2022-10-14T07:12:28.800	&	 $11_{-4}^{+4}$	&	 $67_{-8}^{+8}$	&	 $-14.4_{-3.5}^{+3.4}$	&	 $152.0_{-1.6}^{+1.5}$	&	 $1.6_{-0.7}^{+0.6}$	&	 $250_{-6}^{+7}$	&	 $59.5_{-0.4}^{+0.4}$	&	 $9_{-5}^{+5}$	&	 $126_{-4}^{+4}$ \\

\hline
\end{tabular*}
\label{tab:ercod_par_fit}
\end{table*}

\begin{table*}[htbp]
  \tiny
  \centering
  \caption{Time integrated spectra fitting results of ERCOD and FRED pattern of MXBs.}
  \begin{tabular*}{\textwidth}{ccccccccccccc}
\toprule
 & & & \multicolumn{5}{c}{BB} & \multicolumn{5}{c}{CPL ~(fixed~$\alpha=1$)} \\ 
\cmidrule(r){1-3} \cmidrule(r){4-8} \cmidrule(r){9-13}
Burst ID	&	$\rm T_0$ (UTC)	&   Pattern	& $kT$	&	$fluence$	&	$flux$	&	$\rm \frac{stat}{dof}$	& WAIC & $E_{\rm p}$ & $fluence$ & 	$flux$	&$\rm \frac{stat}{dof}$ & WAIC \\

 & & &  & $\times 10^{-7}$ & $\times 10^{-6}$ & & & & $\times 10^{-7}$ &$\times 10^{-6}$  & & \\
 & & & (keV) & $\mathrm{erg\,cm^{-2}}$ & $\mathrm{erg\,cm^{-2}\,s^{-1}}$ & & & (keV) & $\mathrm{erg\,cm^{-2}}$ & $\mathrm{erg\,cm^{-2}\,s^{-1}}$ & & \\
\hline
018	&	 2021-07-12T04:32:39.600	&	 FRED	&	 $9.6_{-0.5}^{+0.6}$	&	 $0.49_{-0.05}^{+0.05}$	&	 $1.39_{-0.13}^{+0.14}$	&	 1.19	&	 117	&	 $26.5_{-2.6}^{+3.0}$	&	 $0.51_{-0.05}^{+0.05}$	&	 $1.47_{-0.15}^{+0.16}$	&	 1.1	&	 107 \\
019	&	 2021-07-12T22:12:58.100	&	 FRED	&	 $7.2_{-0.4}^{+0.5}$	&	 $0.33_{-0.04}^{+0.04}$	&	 $0.72_{-0.09}^{+0.09}$	&	 1.11	&	 117	&	 $16.3_{-1.7}^{+2.0}$	&	 $0.34_{-0.05}^{+0.05}$	&	 $0.76_{-0.10}^{+0.11}$	&	 1.11	&	 116 \\
026	&	 2021-09-10T03:24:47.150	&	 FRED	&	 $8.29_{-0.26}^{+0.28}$	&	 $1.08_{-0.06}^{+0.07}$	&	 $1.96_{-0.12}^{+0.12}$	&	 1.1	&	 140	&	 $19.4_{-1.1}^{+1.1}$	&	 $1.07_{-0.07}^{+0.07}$	&	 $1.95_{-0.12}^{+0.12}$	&	 1.1	&	 138 \\
029	&	 2021-09-10T05:35:55.500	&	 FRED	&	 $8.7_{-0.7}^{+0.7}$	&	 $0.34_{-0.05}^{+0.05}$	&	 $0.84_{-0.12}^{+0.13}$	&	 1.28	&	 164	&	 $23.0_{-3.2}^{+4.0}$	&	 $0.37_{-0.05}^{+0.06}$	&	 $0.93_{-0.14}^{+0.15}$	&	 1.23	&	 157 \\
033	&	 2021-09-11T16:50:03.850	&	 FRED	&	 $8.02_{-0.21}^{+0.21}$	&	 $1.14_{-0.06}^{+0.06}$	&	 $3.24_{-0.16}^{+0.17}$	&	 1.19	&	 122	&	 $18.0_{-0.8}^{+0.8}$	&	 $1.10_{-0.06}^{+0.06}$	&	 $3.16_{-0.17}^{+0.16}$	&	 1.18	&	 120 \\
047	&	 2021-09-12T00:34:37.450	&	 FRED	&	 $9.37_{-0.14}^{+0.14}$	&	 $4.43_{-0.12}^{+0.12}$	&	 $7.38_{-0.19}^{+0.20}$	&	 1.39	&	 207	&	 $23.3_{-0.6}^{+0.6}$	&	 $4.33_{-0.12}^{+0.13}$	&	 $7.21_{-0.20}^{+0.21}$	&	 1.44	&	 215 \\
067	&	 2021-09-29T23:41:12.245	&	 FRED	&	 $6.7_{-0.5}^{+0.5}$	&	 $0.159_{-0.026}^{+0.029}$	&	 $0.80_{-0.13}^{+0.14}$	&	 1.01	&	 109	&	 $15.3_{-1.9}^{+2.2}$	&	 $0.175_{-0.030}^{+0.032}$	&	 $0.87_{-0.15}^{+0.16}$	&	 0.99	&	 105 \\
077	&	 2022-01-09T07:39:10.700	&	 FRED	&	 $8.2_{-0.4}^{+0.4}$	&	 $0.43_{-0.04}^{+0.04}$	&	 $1.07_{-0.10}^{+0.11}$	&	 1.26	&	 113	&	 $19.4_{-1.6}^{+1.8}$	&	 $0.42_{-0.04}^{+0.05}$	&	 $1.05_{-0.11}^{+0.11}$	&	 1.29	&	 115 \\
091	&	 2022-01-14T19:45:08.100	&	 FRED	&	 $8.0_{-0.6}^{+0.6}$	&	 $0.32_{-0.05}^{+0.05}$	&	 $0.79_{-0.11}^{+0.12}$	&	 1.31	&	 159	&	 $20.3_{-2.5}^{+2.8}$	&	 $0.34_{-0.05}^{+0.05}$	&	 $0.85_{-0.12}^{+0.13}$	&	 1.28	&	 154 \\
115	&	 2022-05-21T23:32:58.800	&	 FRED	&	 $8.0_{-0.4}^{+0.5}$	&	 $0.41_{-0.04}^{+0.05}$	&	 $1.16_{-0.12}^{+0.13}$	&	 1.18	&	 150	&	 $18.8_{-1.8}^{+2.1}$	&	 $0.42_{-0.05}^{+0.05}$	&	 $1.19_{-0.14}^{+0.14}$	&	 1.18	&	 150 \\
117	&	 2022-05-24T17:10:18.550	&	 FRED	&	 $9.5_{-0.4}^{+0.4}$	&	 $0.74_{-0.06}^{+0.06}$	&	 $1.48_{-0.11}^{+0.12}$	&	 1.11	&	 136	&	 $24.8_{-2.0}^{+2.2}$	&	 $0.75_{-0.06}^{+0.06}$	&	 $1.50_{-0.12}^{+0.13}$	&	 1.09	&	 132 \\
132	&	 2022-10-14T11:27:32.750	&	 FRED	&	 $8.5_{-0.4}^{+0.5}$	&	 $0.65_{-0.05}^{+0.05}$	&	 $1.63_{-0.12}^{+0.12}$	&	 1.13	&	 120	&	 $18.0_{-1.5}^{+1.6}$	&	 $0.64_{-0.05}^{+0.05}$	&	 $1.60_{-0.12}^{+0.12}$	&	 1.16	&	 122 \\
140	&	 2022-10-22T01:41:18.050	&	 FRED	&	 $9.5_{-0.8}^{+0.8}$	&	 $0.40_{-0.04}^{+0.04}$	&	 $1.59_{-0.14}^{+0.15}$	&	 1.08	&	 93	&	 $19.2_{-2.4}^{+2.8}$	&	 $0.40_{-0.04}^{+0.04}$	&	 $1.62_{-0.15}^{+0.16}$	&	 1.07	&	 91 \\
006	&	 2021-01-27T06:50:20.750	&	 ERCOD	&	 $9.02_{-0.13}^{+0.14}$	&	 $7.06_{-0.17}^{+0.17}$	&	 $7.06_{-0.17}^{+0.17}$	&	 1.73	&	 137	&	 $20.0_{-0.5}^{+0.5}$	&	 $6.88_{-0.17}^{+0.17}$	&	 $6.88_{-0.17}^{+0.17}$	&	 1.1	&	 86 \\
011	&	 2021-01-30T17:40:54.750	&	 ERCOD	&	 $11.85_{-0.13}^{+0.13}$	&	 $15.66_{-0.26}^{+0.26}$	&	 $14.92_{-0.24}^{+0.25}$	&	 2.99	&	 289	&	 $31.1_{-0.6}^{+0.7}$	&	 $15.46_{-0.27}^{+0.29}$	&	 $14.72_{-0.26}^{+0.27}$	&	 1.51	&	 144 \\
014	&	 2021-02-16T22:20:39.600	&	 ERCOD	&	 $8.42_{-0.12}^{+0.12}$	&	 $7.10_{-0.16}^{+0.16}$	&	 $7.48_{-0.17}^{+0.17}$	&	 1.74	&	 163	&	 $17.9_{-0.4}^{+0.4}$	&	 $6.92_{-0.16}^{+0.16}$	&	 $7.29_{-0.17}^{+0.17}$	&	 1.6	&	 150 \\
015	&	 2021-07-07T00:33:31.640	&	 ERCOD	&	 $9.89_{-0.12}^{+0.12}$	&	 $11.18_{-0.20}^{+0.20}$	&	 $10.65_{-0.19}^{+0.19}$	&	 1.25	&	 132	&	 $21.9_{-0.5}^{+0.5}$	&	 $10.89_{-0.20}^{+0.20}$	&	 $10.37_{-0.19}^{+0.19}$	&	 1.27	&	 133 \\
030	&	 2021-09-11T05:32:38.620	&	 ERCOD	&	 $8.27_{-0.18}^{+0.18}$	&	 $2.84_{-0.09}^{+0.09}$	&	 $5.16_{-0.16}^{+0.16}$	&	 1.08	&	 130	&	 $16.6_{-0.6}^{+0.6}$	&	 $2.77_{-0.09}^{+0.09}$	&	 $5.03_{-0.17}^{+0.17}$	&	 1.06	&	 127 \\
034	&	 2021-09-11T17:01:10.800	&	 ERCOD	&	 $9.55_{-0.05}^{+0.05}$	&	 $67.8_{-0.5}^{+0.5}$	&	 $12.56_{-0.09}^{+0.09}$	&	 2.17	&	 281	&	 $20.84_{-0.17}^{+0.17}$	&	 $66.4_{-0.5}^{+0.5}$	&	 $12.29_{-0.09}^{+0.09}$	&	 2.63	&	 343 \\
106	&	 2022-01-15T17:21:59.300	&	 ERCOD	&	 $10.82_{-0.12}^{+0.12}$	&	 $13.07_{-0.21}^{+0.21}$	&	 $9.68_{-0.15}^{+0.16}$	&	 1.78	&	 207	&	 $26.0_{-0.5}^{+0.5}$	&	 $12.93_{-0.21}^{+0.22}$	&	 $9.58_{-0.15}^{+0.16}$	&	 1.22	&	 142 \\
111	&	 2022-01-23T20:06:38.750	&	 ERCOD	&	 $8.99_{-0.06}^{+0.06}$	&	 $30.14_{-0.29}^{+0.31}$	&	 $10.05_{-0.10}^{+0.10}$	&	 3.08	&	 298	&	 $19.17_{-0.22}^{+0.22}$	&	 $29.27_{-0.31}^{+0.31}$	&	 $9.76_{-0.10}^{+0.10}$	&	 3.92	&	 379 \\
126	&	 2022-10-12T15:45:10.150	&	 ERCOD	&	 $12.16_{-0.12}^{+0.12}$	&	 $22.33_{-0.33}^{+0.33}$	&	 $12.07_{-0.18}^{+0.18}$	&	 4.77	&	 401	&	 $30.0_{-0.5}^{+0.6}$	&	 $21.86_{-0.33}^{+0.34}$	&	 $11.81_{-0.18}^{+0.18}$	&	 9.23	&	 774 \\
130	&	 2022-10-14T07:12:28.800	&	 ERCOD	&	 $9.11_{-0.13}^{+0.13}$	&	 $8.46_{-0.17}^{+0.17}$	&	 $5.46_{-0.11}^{+0.11}$	&	 1.2	&	 163	&	 $19.9_{-0.5}^{+0.5}$	&	 $8.41_{-0.17}^{+0.18}$	&	 $5.42_{-0.11}^{+0.11}$	&	 1.04	&	 142 \\
\hline
\\
\end{tabular*}
\label{tab:spec_fitting_result}
\footnotesize
\textbf{NOTE}: 
(1) The Blackbody model is defined 
$N(E) = \frac{C K E^2}{(kT)^4 [\exp(E/kT)-1]}$,\\
\bigskip
(2) The CPL model is defined as $N(E) = K \left(\frac{E}{E_0}\right)^{-\alpha}\exp \left[-\frac{(2-\alpha)E}{E_\mathrm{p}}\right]$, where $E_0$ is the reference energy fixed at 1 keV,\\
\bigskip
(3) The calculated fluence and flux in the table is between 35-300 keV.\\
\end{table*}

\begin{table*}[htbp]
\caption{\centering{Time, energy range and detectors selection for spectral analysis.}}
\tiny
\begin{tabular*}{\hsize}
{@{}@{\extracolsep{\fill}}cccccccc@{}}
\hline
\toprule
Burst ID	&	$\rm T_0$ (UTC)	&	Pattern	&	$T_{\rm90}~\rm(ms)$ &	$T_{\rm90}~\rm start~(ms)$	&	$T_{\rm90}~\rm end~(ms)$&	Energy range (keV)  & GRD  \\

\hline
18	&	 2021-07-12T04:32:39.600	&	 FRED	&	 $35^{+15}_{-10}$	&	-43.21	&	-3.21	 & 	 [26,300]	&	 ['12', '13', '14', '21', '22', '23', '05'] \\
19	&	 2021-07-12T22:12:58.100	&	 FRED	&	 $45^{+10}_{-15}$	&	-64.60	&	-19.60	 & 	 [26,300]	&	 ['12', '13', '14', '21', '22', '23', '05'] \\
26	&	 2021-09-10T03:24:47.150	&	 FRED	&	 $55^{+20}_{-10}$	&	-19.11	&	30.89	 & 	 [26,300]	&	 ['12', '13', '14', '15', '22', '23', '05', '06'] \\
29	&	 2021-09-10T05:35:55.500	&	 FRED	&	 $40^{+10}_{-15}$	&	-24.96	&	10.04	 & 	 [26,300]	&	 ['12', '13', '14', '15', '22', '23', '05', '06'] \\
33	&	 2021-09-11T16:50:03.850	&	 FRED	&	 $35^{+5}_{-5}$	&	-24.79	&	10.21	     & 	 [26,300]	&	 ['12', '13', '14', '15', '22', '23', '05', '06'] \\
47	&	 2021-09-12T00:34:37.450	&	 FRED	&	 $60^{+5}_{-5}$	&	-43.73	&	16.27	     & 	 [26,300]	&	 ['12', '13', '14', '15', '22', '23', '05', '06'] \\
67	&	 2021-09-29T23:41:12.245	&	 FRED	&	 $20^{+15}_{-5}$	&	-9.69	&	10.31	 & 	 [25,300]	&	 ['10', '11', '12', '20', '21', '03', '04'] \\
77	&	 2022-01-09T07:39:10.700	&	 FRED	&	 $40^{+20}_{-15}$	&	-24.78	&	10.22	 & 	 [24,300]	&	 ['15', '16', '17', '01', '02', '05', '06', '07'] \\
91	&	 2022-01-14T19:45:08.100	&	 FRED	&	 $40^{+30}_{-25}$	&	-38.23	&	-8.23	 & 	 [24,300]	&	 ['15', '16', '17', '01', '02', '05', '06', '07'] \\
115	&	 2022-05-21T23:32:58.800	&	 FRED	&	 $35^{+15}_{-10}$	&	-32.31	&	-2.31	 & 	 [26,300]	&	 ['13', '14', '15', '22', '23', '24', '05', '06'] \\
117	&	 2022-05-24T17:10:18.550	&	 FRED	&	 $50^{+15}_{-20}$	&	-35.00	&	10.00	 & 	 [26,300]	&	 ['13', '14', '15', '22', '23', '24', '05', '06'] \\
132	&	 2022-10-14T11:27:32.750	&	 FRED	&	 $40^{+15}_{-20}$	&	-34.14	&	0.86	 & 	 [32,300]	&	 ['11', '12', '13', '20', '21', '22', '03', '04', '05'] \\
140	&	 2022-10-22T01:41:18.050	&	 FRED	&	 $25^{+20}_{-5}$	&	-53.74	&	-28.74	 &	 [41,300]	&	 ['10', '17', '18', '19', '02', '03', '07', '08', '09'] \\
6	&	 2021-01-27T06:50:20.750	&	 ERCOD	&	 $100^{+25}_{-0}$	&	-38.71	&	63.79	 & 	 [32,300]	&	 ['16', '17', '18', '25', '07', '08'] \\
11	&	 2021-01-30T17:40:54.750	&	 ERCOD  &	 $105^{+0}_{-5}$	&	1.65	&	106.65	 & 	 [32,300]	&	 ['16', '17', '18', '25', '07', '08'] \\
14	&	 2021-02-16T22:20:39.600	&	 ERCOD	&	 $95^{+5}_{-0}$	&	205.06	&	300.06	     & 	 [32,300]	&	 ['16', '17', '18', '24', '25', '07', '08'] \\
15	&	 2021-07-07T00:33:31.640	&	 ERCOD	&	 $105^{+5}_{-5}$	&	11.50	&	116.50	 & 	 [32,300]	&	 ['12', '13', '14', '21', '22', '23', '05'] \\
30	&	 2021-09-11T05:32:38.620	&	 ERCOD	&	 $55^{+5}_{-0}$	&	21.14	&	76.14	     & 	 [32,300]	&	 ['12', '13', '14', '15', '22', '23', '05', '06'] \\
34	&	 2021-09-11T17:01:10.800	&	 ERCOD	&	 $540^{+0}_{-5}$	&	-324.31	&	215.69	 & 	 [32,300]	&	 ['12', '13', '14', '15', '22', '23', '05', '06'] \\
106	&	 2022-01-15T17:21:59.300	&	 ERCOD	&	 $135^{+20}_{-0}$	&	-13.89	&	121.11	 & 	 [32,300]	&	 ['15', '16', '17', '01', '02', '05', '06', '07'] \\
111	&	 2022-01-23T20:06:38.750	&	 ERCOD	&	 $300^{+0}_{-5}$	&	1.12	&	301.12	 & 	 [32,300]	&	 ['15', '16', '17', '01', '02', '05', '06', '07'] \\
126	&	 2022-10-12T15:45:10.150	&	 ERCOD	&	 $185^{+25}_{-10}$	&	-168.61	&	11.39	 & 	 [32,300]	&	 ['14', '15', '16', '23', '24', '25'] \\
130	&	 2022-10-14T07:12:28.800	&	 ERCOD	&	 $155^{+15}_{-10}$	&	-29.86	&	125.14	 & 	 [33,300]	&	 ['10', '11', '19', '20', '21', '03', '04', '09'] \\
\hline
\end{tabular*}
\label{tab:t90_energy_grd}
\end{table*}


\clearpage
\bibliography{main}
\bibliographystyle{aasjournal}

\end{document}